\documentclass[preprint,preprintnumbers,showpacs,aps,prd,amssymb]{revtex4-1}

\usepackage{graphicx}
\usepackage{bm}
\usepackage{amsmath}

\def\calH{{\cal H}}

\def\calO{{\cal O}}

\def\calU{{\cal U}}

\def\Bbar{{\bar B}}

\def\hbar{{\bar h}}

\def\sbar{{\bar s}}

\def\SM{{\rm SM}}
\def\Br{{\rm Br}}

\def\Ks{K^{(*)}} 
\def\mhat{{\hat m}}
\def\shat{{\hat s}}
\def\uhat{{\hat u}}

\def\Bs2mumu{{B_s\to\mu^+\mu^-}}

\def\nn{\nonumber}

\begin{document}
\title{Model-independent analysis on $R(K^{(*)})$}
\author{Jong-Phil Lee}
\email{jongphil7@gmail.com}
\affiliation{Sang-Huh College,
Konkuk University, Seoul 05029, Korea}

\begin{abstract}
We analyze the lepton-universality violating $B\to K^{(*)}\ell^+\ell^-$ puzzle $R(K^{(*)})$ in a model-independent way.
The branching ratio ${\rm Br}(B_s\to\mu^+\mu^-)$ is also considered as a main constraint.
We show how the branching ratio restricts the allowed region of the parameter space,
and investigate new physics effects on the electronic as well as muonic sector in $R(K^{(*)})$.
Within a reasonable range of parameters we find that the new physics scale goes up to $\sim 5$ TeV,
and the new physics effects on the Wilson coefficient is $-1\lesssim C_{9NP}^\mu\lesssim 0$.
\end{abstract}

\maketitle
\section{Introduction}
Flavor physics has played an important role to test the standard model (SM) and to probe new physics (NP).
Various experimental results are in good agreement with the SM but there are some anomalies in $B$ meson phenomenology.
Recently some hints for the lepton-universality violation in the semileptonic $B$ decays are reported.
One of them is the so called $R(\Ks)$, which is the ratio of the branching ratio of $B\to\Ks\ell\ell$ decays, defined by
\begin{equation}
R(\Ks)\equiv\frac{\Br(B\to\Ks\mu^+\mu^-)}{\Br(B\to\Ks e^+e^-)}~.
\end{equation}
The process is a kind of the flavor-changing neutral current (FCNC) which is not allowed at the tree level in the SM.
One can expect some NP effects could explain the involved anomalies.
Recently updated measurements from the LHCb Collaboration are \cite{LHCb2103,LHCb1705,Geng2103},
\begin{eqnarray}
R(K)[1.1,6.0] &=& 0.846^{+0.044}_{-0.041}~,  \nn\\
R(K^*)[0.045,1.1] &=& 0.660^{+0.110}_{-0.070}\pm0.024~, \nn\\
R(K^*)[1.1,6.0] &=& 0.685^{+0.113}_{-0.069}\pm0.047~,
\label{RKs_EXP}
\end{eqnarray}
where the numbers in the square bracket are the bins of the momentum squared in ${\rm GeV}^2$.
The results are in tension with the SM predictions which are quite close to unity \cite{Hiller0310,Bobeth0709,Geng1704},
\begin{eqnarray}
R(K)_\SM[1.0,6.0] &=&1.0004^{+0.0008}_{-0.0007}~, \nn\\
R(K^*)_\SM[0.045,1.1] &=& 0.920^{+0.007}_{-0.006}~, \nn\\
R(K^*)_\SM[1.1,6.0] &=& 0.996^{+0.002}_{-0.002}~.
\label{RKs_SM}
\end{eqnarray}
The discrepancy between the experiments and the SM strongly suggests a need for NP to explain the 
lepton-universality violation.
Up to now many NP scenarios have been proposed to solve the puzzle, including 
the leptoquark \cite{Hiller1408,Dorsner1603,Bauer1511,Chen1703,Crivellin1703,Calibbi1709,Blanke1801,Nomura2104,Angelescu2103,Du2104}, 
$Z'$ model \cite{Crivellin1501,Crivellin1503,Chiang1706,King1706,Chivukula1706,Cen2104,Davighi2105}, 
two-Higgs doublet model \cite{Hu1612,Crivellin1903,Rose1903}.
and so on \cite{Altmannshofer2002}.
\par
In this paper, we analyze the $R(\Ks)$ puzzle in a model-independent way.
This kind of approach makes it possible to show general features of the situation.
The NP effects are generally encoded in the relevant Wilson coefficients $C_i$.
In many cases NP involves a new scale $M_{NP}$ in the form of $\sim 1/(G_F M_{NP}^2)$.
We parameterize the Wilson coefficient from NP as $C_{iNP}\sim (v/M_{NP})^\alpha$ with some coefficients, 
where $v$ is the SM vacuum expectation value.
Usually $\alpha=2$ but in some NP like unparticle scenario $\alpha$ could be free parameters \cite{Georgi,JPL2106}.
As will be seen later, our simple parametrization can provide some information about the NP scale $M_{NP}$
and different patterns between electronic and muonic contributions.
\par
In the next Section we provide our formalism to do the analysis.
Section III contains the results and discussions.
We conclude in Sec.\ IV.
%
\section{Formalism}
%
The $b\to s\ell^+\ell^-$ transition is described by the effective Hamiltonian
\begin{equation}
\calH_{\rm eff} = -\frac{4G_F}{\sqrt{2}}V_{tb}V_{ts}^*\sum_{i=9}^{10} C_i(\mu)\calO_i(\mu)~.
\end{equation} 
For the $R(\Ks)$ anomaly It is well known that only $\calO_9$ and $\calO_{10}$ operators are relevant
as discussed in \cite{Alonso14,Geng17,Geng21}
\begin{eqnarray}
\calO_9 &=& \frac{e^2}{16\pi^2}\left(\sbar\gamma^\mu P_L b\right)\left({\bar\ell}\gamma_\mu\ell\right)~,\nn\\
\calO_{10} &=& \frac{e^2}{16\pi^2}\left(\sbar\gamma^\mu P_L b\right)\left({\bar\ell}\gamma_\mu\gamma_5\ell\right)~.
\label{O9O10}
\end{eqnarray}
The matrix elements for $B\to\Ks$ are given by ($q=p_B-p$) \cite{Ali99}
\begin{eqnarray}
\langle K(p)|\sbar\gamma_\mu b|B(p_B)\rangle&=&
f_+\left[(p_B+p)_\mu-\frac{m_B^2-m_K^2}{s}q_\mu\right]+\frac{m_B^2-m_K^2}{s}f_0 q_\mu~,\\
\langle K(p)|\sbar\sigma_{\mu\nu} q^\nu(1+\gamma_5)b|B(p_B)\rangle&=&
i\left[(p_B+p)_\mu s -q_\mu(m_B^2-m_K^2)\right]\frac{f_T}{m_B+m_K}~,\\
\langle K^*(p)|(V-A)_\mu|B(p_B)\rangle&=&
-i\epsilon_\mu^*(m_B+m_{K^*})A_1+i(p_B+p)_\mu(\epsilon^*\cdot p_B)\frac{A_2}{m_B+m_{K^*}}\nn\\
&&
+iq_\mu(\epsilon^*\cdot p_B)\frac{2m_{K^*}}{s}(A_3-A_0)
+\frac{\epsilon_{\mu\nu\rho\sigma}\epsilon^{*\nu}p_B^\rho b^\sigma}{m_B+m_{K^*}}2V~,
\end{eqnarray}
where $f_{+,0,T}(s), ~A_{0,1,2}(s), ~T_{1,2,3}(s), V(s)$, and $f_-=(f_0-f_+)(1-\mhat_K^2)/\shat$ are the form factors.
We adopt the exponential forms of \cite{Ali99} for the form factors. 
\par
Now the differential decay rates for $B\to\Ks\ell^+\ell^-$ are \cite{Chang2010}
\begin{eqnarray}
\frac{d\Gamma_K}{d\shat}&=&
\frac{G_F^2\alpha^2m_B^5}{2^{10}\pi^5}|V_{tb}V_{ts}^*|^2\uhat_{K,\ell}\left\{
(|A'|^2+|C'|^2)\left(\lambda_K-\frac{\uhat_{K,\ell}^2}{3}\right)
+|C'|^24\mhat_\ell^2(2+2\mhat_K^2-\shat) \right .\nn\\
&&\left. +{\rm Re}(C'D'^*)8\mhat_\ell^2(1-\mhat_K^2)+|D'|^24\mhat_\ell^2\shat\right\}~,
\\
\frac{d\Gamma_{K^*}}{d\shat}&=&
\frac{G_F^2\alpha^2m_B^5}{2^{10}\pi^5}|V_{tb}V_{ts}^*|^2\uhat_{K^*,\ell}\left\{
\frac{|A|^2}{3}\shat\lambda_{K^*}\left(1+\frac{2\mhat_\ell^2}{\shat}\right)
+|E|^2\shat\frac{\uhat_{K^*,\ell}^2}{3}\right.\nn\\
&&
+\frac{|B|^2}{4\mhat_{K^*}^2}\left[\lambda_{K^*}-\frac{\uhat_{K^*,\ell}t^2}{3}+8\mhat^2_{K^*}(\shat+2\mhat_\ell^2)\right]
+\frac{|F|^2}{4\mhat_{K^*}^2}\left[\lambda_{K^*}-\frac{\uhat_{K^*,\ell}^2}{3}+8\mhat_{K^*}^2(\shat-4\mhat_\ell^2)\right]\nn\\
&&
+\frac{\lambda_{K^*}|C|^2}{4\mhat_{K^*}^2}\left(\lambda_{K^*}-\frac{\uhat_{K^*,\ell}^2}{3}\right)
+\frac{\lambda|_{K^*}|G|^2}{4\mhat_{K^*}^2}\left[\lambda_{K^*}-\frac{\uhat_{K^*,\ell}^2}{3}
	+4\mhat_\ell^2(2+2\mhat_{K^*}^2-\shat)\right]\nn\\
&&
-\frac{{\rm Re}(BC^*)}{2\mhat_{K^*}^2}\left(\lambda_{K^*}-\frac{\uhat_{K^*,\ell}^2}{3}\right)(1-\mhat_{K^*}^2-\shat)\nn\\
&&
-\frac{{\rm Re}(FG^*)}{2\mhat_{K^*}^2}\left[\left(\lambda_{K^*}-\frac{\uhat_{K^*,\ell}^2}{3}\right)(1-\mhat_{K^*}^2
	-\shat)-4\mhat_\ell^2\lambda_{K^*}\right]\nn\\
&&\left.
-\frac{2\mhat_\ell^2}{\mhat_{K^*}^2}\lambda_{K^*}\left[{\rm Re}(FH^*)-{\rm Re}(GH^*)(1-\mhat_{K^*}^2)\right]
+\frac{\mhat_\ell^2}{\mhat_{K^*}^2}\shat\lambda_{K^*}|H|^2\right\}~,
\end{eqnarray}
where the kinematic variables are
\begin{eqnarray}
\shat&=&\frac{s}{m_B^2}~,~~~\mhat_i=\frac{m_i}{m_B}~,\\
\lambda_H&=&1+\mhat_H^4+\shat^2-2\shat-2\mhat_H^2(1+\shat)~,~~~
\uhat_{H,\ell}=\sqrt{\lambda_H\left(1-\frac{4\mhat_\ell^2}{\shat}\right)}~,
\end{eqnarray}
and
\begin{equation}
s=(p_B-p_{\Ks})^2=(p_{\ell^+}+p_{\ell^-})^2~,
\end{equation}
is the momentum transfer squared.
Here the auxiliary functions $A',\cdots, D'$ and $A,\cdots, H$ are defined by the Wilson coefficients and the form factors \cite{Ali99}.
%
\begin{eqnarray}
A'&=&C_9 f_+ +\frac{2\mhat_b}{1+\mhat_K}C_7 f_T ~,\\
B'&=&C_9 f_- -\frac{2\mhat_b}{\shat}(1-\mhat_K)C_7 f_T ~,\\
C'&=&C_{10} f_+ ~,\\
D'&=&C_{10} f_- ~,
\end{eqnarray}
and
\begin{eqnarray}
A&=&\frac{2}{1+\mhat_{K^*}}C_9 V+\frac{4\mhat_b}{\shat}C_7 T_1~,\\
B&=&(1+\mhat_{K^*})\left[C_9 A_1+\frac{2\mhat_b}{\shat}(1-\mhat_{K^*})C_7 T_2\right]~,\\ 
C&=&\frac{1}{1-\mhat_{K^*}^2}\left[(1-\mhat_{K^*})C_9 A_2
	+2\mhat_b C_7 \left(T_3+\frac{1-\mhat_{K^*}^2}{\shat}T_2\right)\right]~,\\
D&=&\frac{1}{\shat}\left\{C_9\left[(1+\mhat_{K^*})A_1-(1-\mhat_{K^*})A_2
	-2\mhat_{K^*}A_0\right]-2\mhat_b C_7 T_3\right\}~,\\
E&=&\frac{2}{1+\mhat_{K^*}}C_{10} V ~,\\
F&=&(1+\mhat_{K^*})C_{10} A_1 ~,\\
G&=&\frac{1}{1+\mhat_{K^*}}C_{10} A_2 ~,\\
H&=&\frac{1}{\shat}C_{10}\left[(1+\mhat_{K^*})A_1-(1-\mhat_{K^*})A_2-2\mhat_{K^*} A_0\right] ~.
\end{eqnarray}
%
\par
Now we parameterize the Wilson coefficient as
\begin{equation}
C_{9NP}^\ell = A_9^\ell\left(\frac{v}{M_{NP}}\right)^\alpha = -C_{10NP}^\ell~,
\end{equation}
where $\ell=e,\mu$.
We consider the power of $(v/M_{NP})$ as a free parameter just like the unparticles.
Usually unparticles contribute in the form of $\sim(v^2/\Lambda_\calU^2)^{d_\calU}$, 
where $\Lambda_\calU\gtrsim 1~{\rm TeV}$ is the unparticle scale and $d_\calU$ is the scaling dimension of the unparticle operator.
Because of the scale-invariant nature of the unparticle $d_\calU$ need not be integers.
\par
As an example, for the leptoquark model of \cite{Du2104}
\begin{equation}
A_{9LQ}^\mu = -\frac{\pi x_{s\mu} x_{b\mu}^*}{\alpha_{em}V_{tb}V_{ts}^*}~,
\end{equation}
where $x_{ij}$ are the relevant couplings.
For the values of $x_{s\mu}=-0.005$ and $x_{b\mu}=0.5$, one gets $A_9^\mu\sim -25$.
If $(v/M_{NP})\sim 10$, then $C_{9NP}\sim -0.25$.
Similarly for new $Z'$ contribution \cite{Cen2104}
\begin{equation}
A_{9Z'}^\mu=\frac{2\pi{\tilde g}^2 V_{L22}^d V_{L32}^{d*}}{3\alpha_{em}V_{tb}V_{ts}^*}~,
\end{equation}
where ${\tilde g}, V_{Lij}^d$ are model parameters.
\par
Constraints for the parameters mainly come from $B_s\to\mu^+\mu^-$ decay.
The NP effects appear in the branching ratio of the $B_s\to\mu^+\mu^-$ as
\begin{equation}
\Br(B_s\to\mu^+\mu^-)_\calU 
= \Br(B_s\to\mu^+\mu^-)_\SM \left|1
	+\frac{C_{10}^{NP}(B_s\to\mu^+\mu^-)}{C_{10}^\SM(B_s\to\mu^+\mu^-)}\right|^2~,
\end{equation}
where the SM prediction is \cite{Geng2103,Beneke1908}
\begin{equation}
\Br(\Bs2mumu)_\SM = (3.63\pm0.13)\times 10^{-9}~,
\label{BrBsmumuSM}
\end{equation}
which must be compared with the experimental result \cite{Geng2103}
\begin{equation}
\Br(B_s\to\mu^+\mu^-)_{\rm exp} = (2.842\pm0.333)\times 10^{-9}~.
\label{BrBsmumuEXP}
\end{equation}
\par
%
\section{Results and Discussions}
%
In this analysis we confine $|A_9^{e,\mu}|\le 100$.
The reason is that in this range for $\alpha=2$ and  $M_{NP}\sim 2~{\rm TeV}$ the 
Wilson coefficients become $|C_{9NP}^\ell|\lesssim\calO(1)$,
which is compatible with other researches like \cite{Geng2103}.
\par
\begin{figure}
\begin{tabular}{cc}
\hspace{-1cm}\includegraphics[scale=0.15]{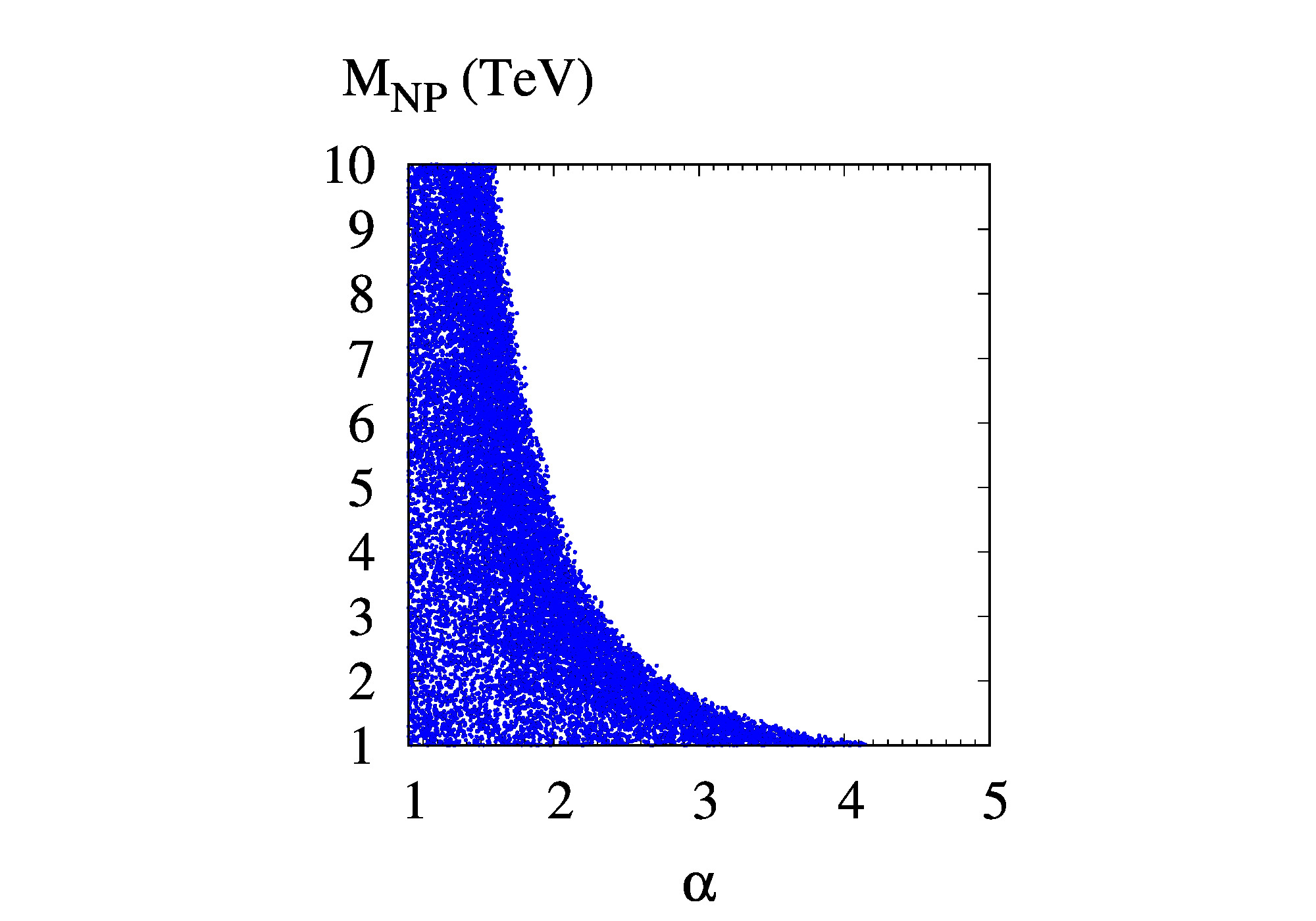} &
\hspace{-2cm}\includegraphics[scale=0.15]{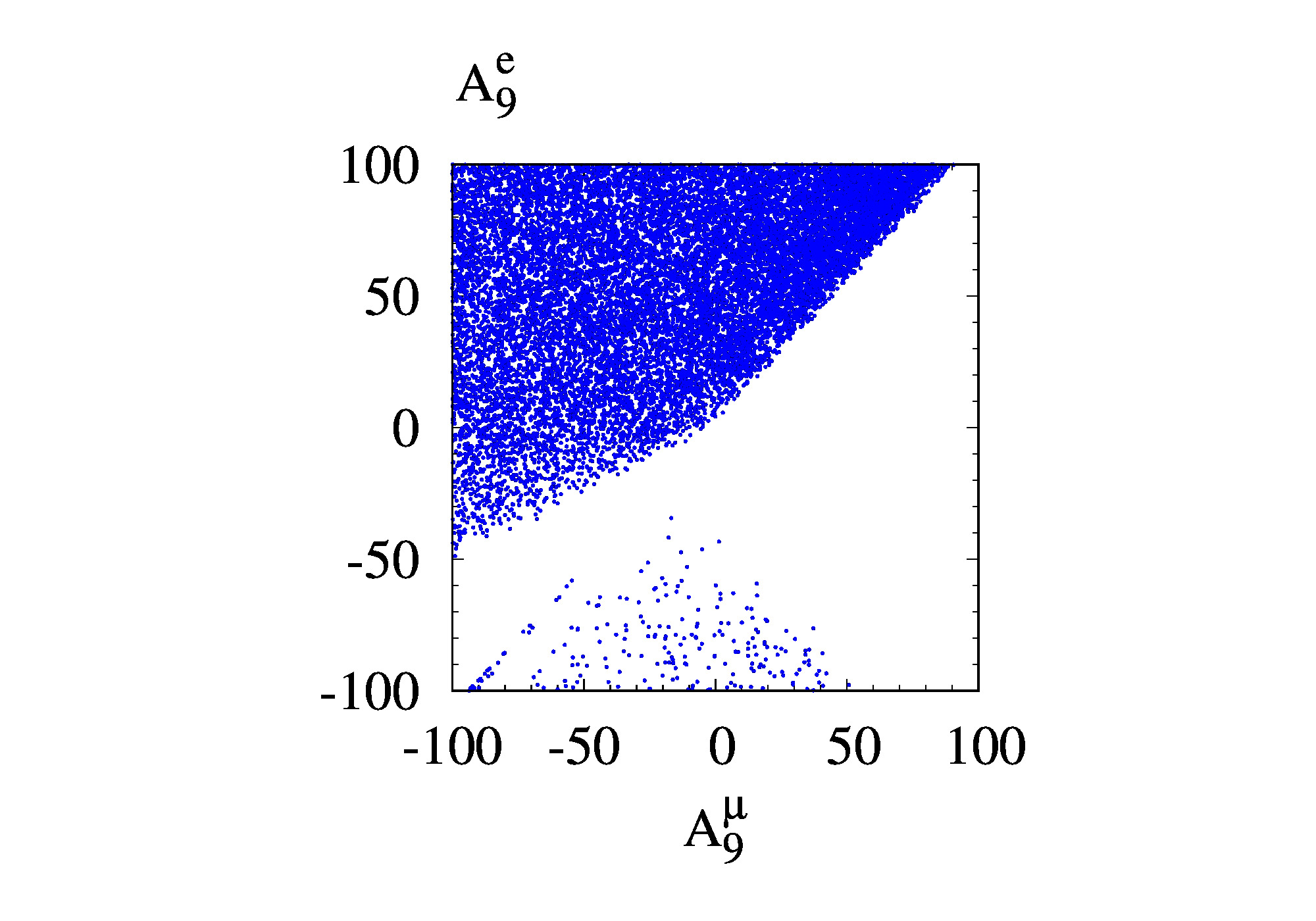}\\
\hspace{-2cm} (a) & \hspace{-1cm} (b) \\
\hspace{-1cm}\includegraphics[scale=0.15]{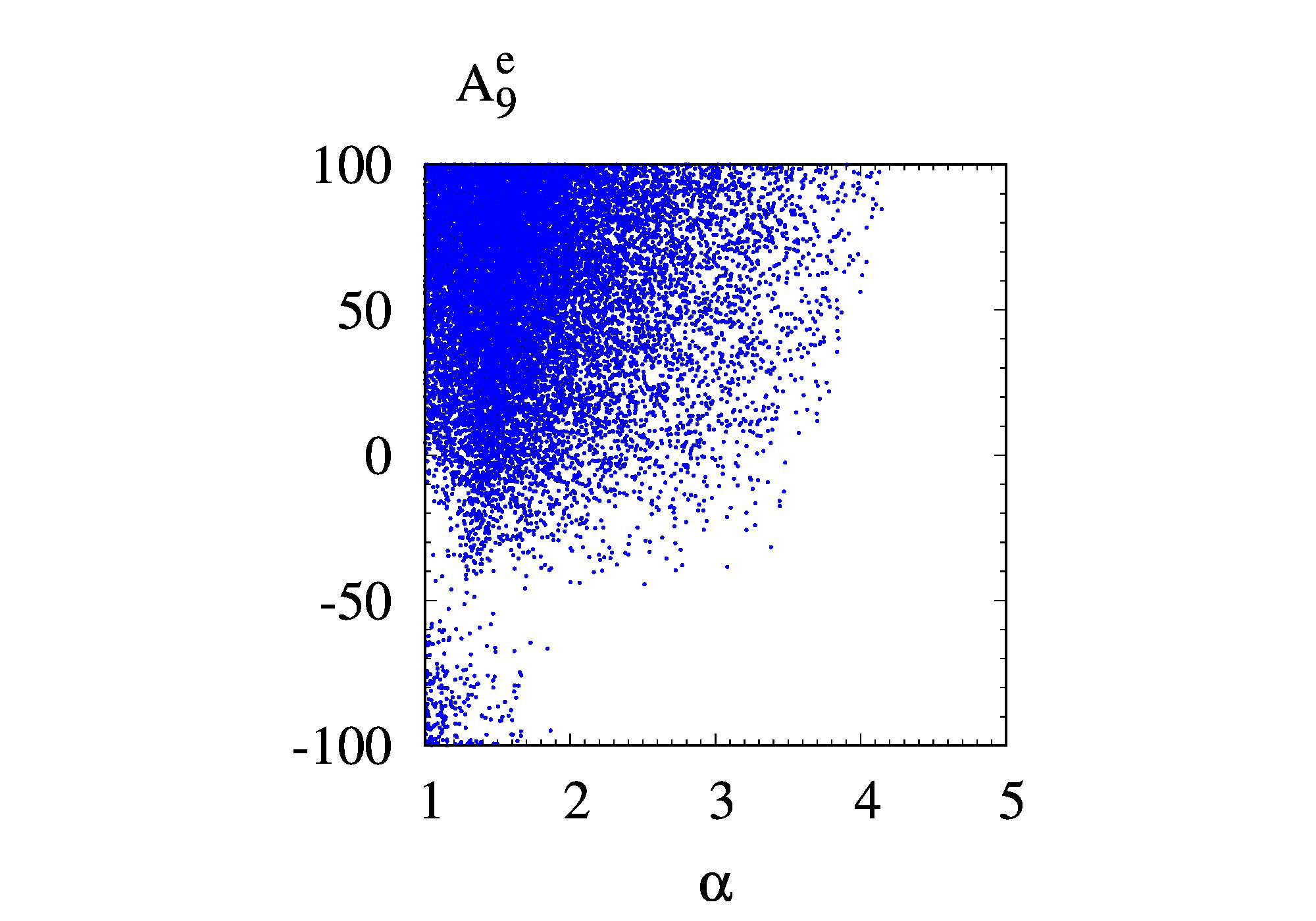} &
\hspace{-2cm}\includegraphics[scale=0.15]{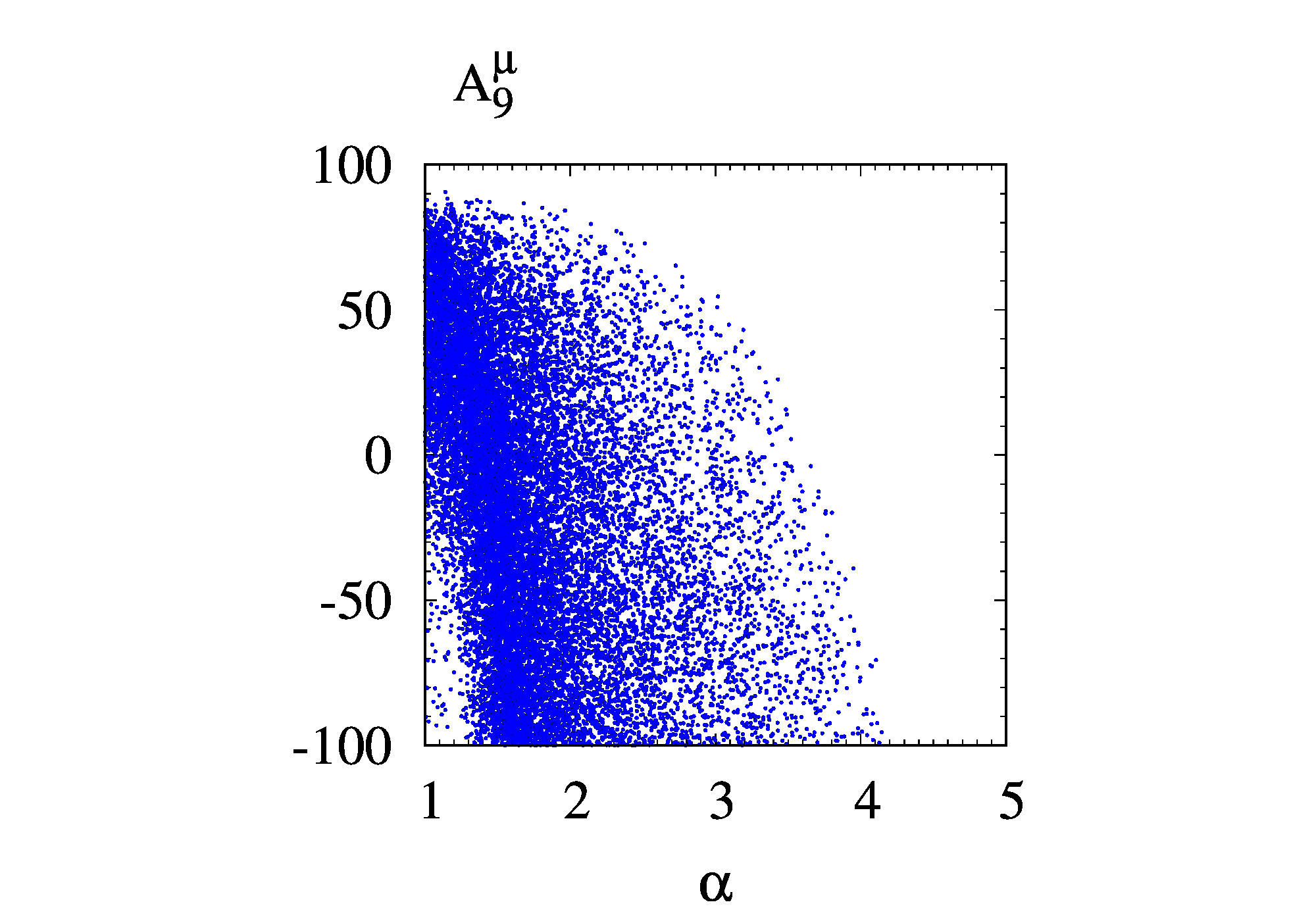}\\
\hspace{-2cm} (c) & \hspace{-1cm} (d) 
\end{tabular}
\caption{\label{F1} Allowed regions of the relevant parameters, (a) $M_{NP}$ vs $\alpha$, (b) $A_9^e$ vs $A_9^\mu$,
(c) $A_9^e$ vs $\alpha$, and (d) $A_9^\mu$ vs $\alpha$ at the $2\sigma$ level without any constraints.}
\end{figure}
\begin{figure}
\begin{tabular}{c}
\hspace{-1cm}\includegraphics[scale=0.15]{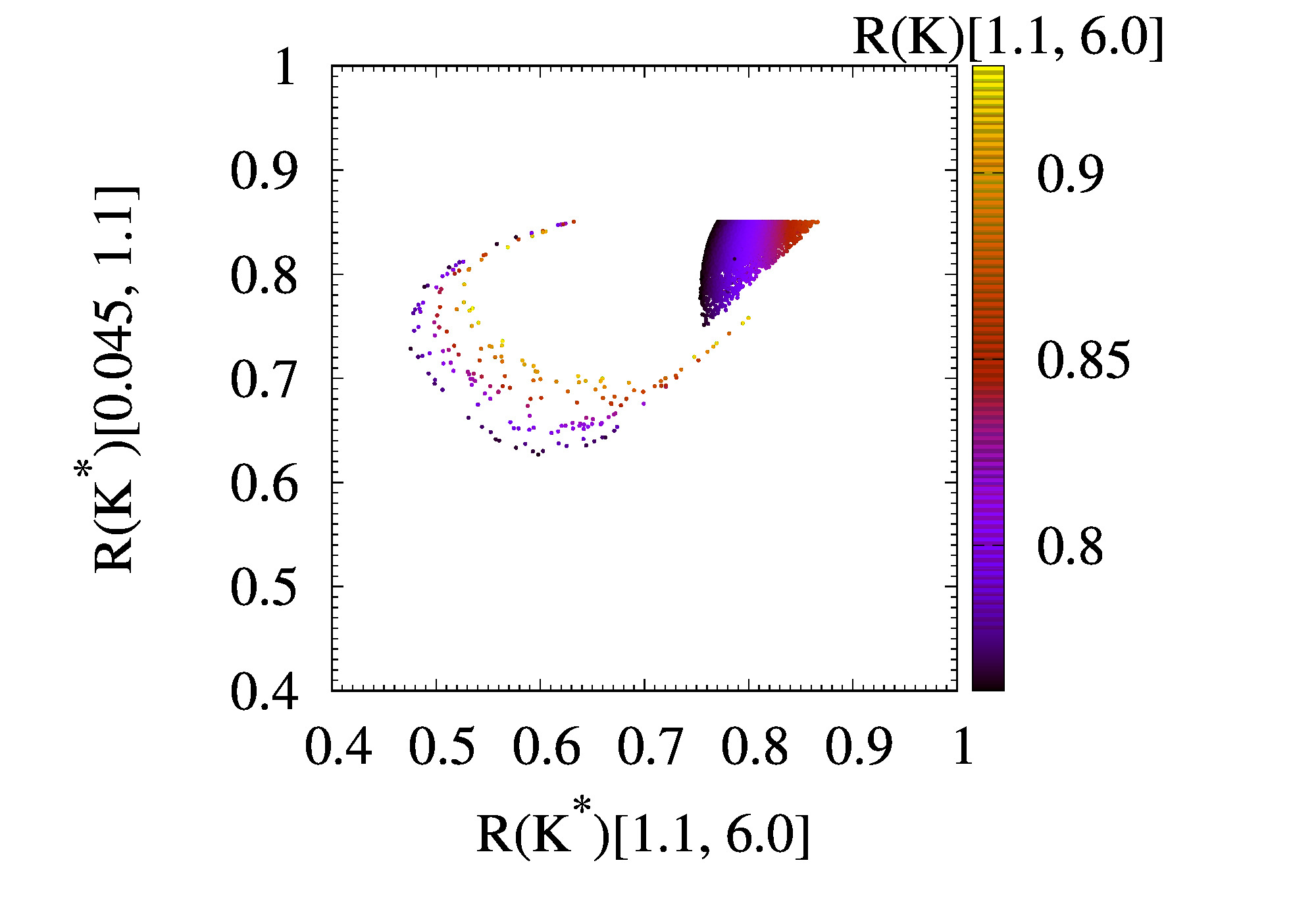} 
\end{tabular}
\caption{\label{no_cst_obs} Allowed regions of $R(K^*)[0.045,1.1]$ vs $R(K^*)[1.1,6.0]$ with respect to
$R(K)[1.1,6.0]$ at the $2\sigma$ level without any constraints.}
\end{figure}

First we do not consider the constraints from $\Br(\Bs2mumu)$.
Figure \ref{F1} shows the allowed region of the relevant parameters for $R(\Ks)$.
The behavior of $M_{NP}$ vs $\alpha$ in Fig.\ \ref{F1} is very common in the unparticle scenario.
Heavier mass of $M_{NP}$ is not allowed as $\alpha$ gets larger since in that case the NP effects would be very suppressed.
Figures \ref{F1} (b)-(d) show that the allowed regions of $A_9^e$ are quite different from those of $A_9^\mu$, 
which explains the lepton flavor violation in $R(\Ks)$.
We plot the distribution of $R(\Ks)$ in our scheme in Fig.\ \ref{no_cst_obs}.
\par
\begin{figure}
\begin{tabular}{cc}
\hspace{-1cm}\includegraphics[scale=0.15]{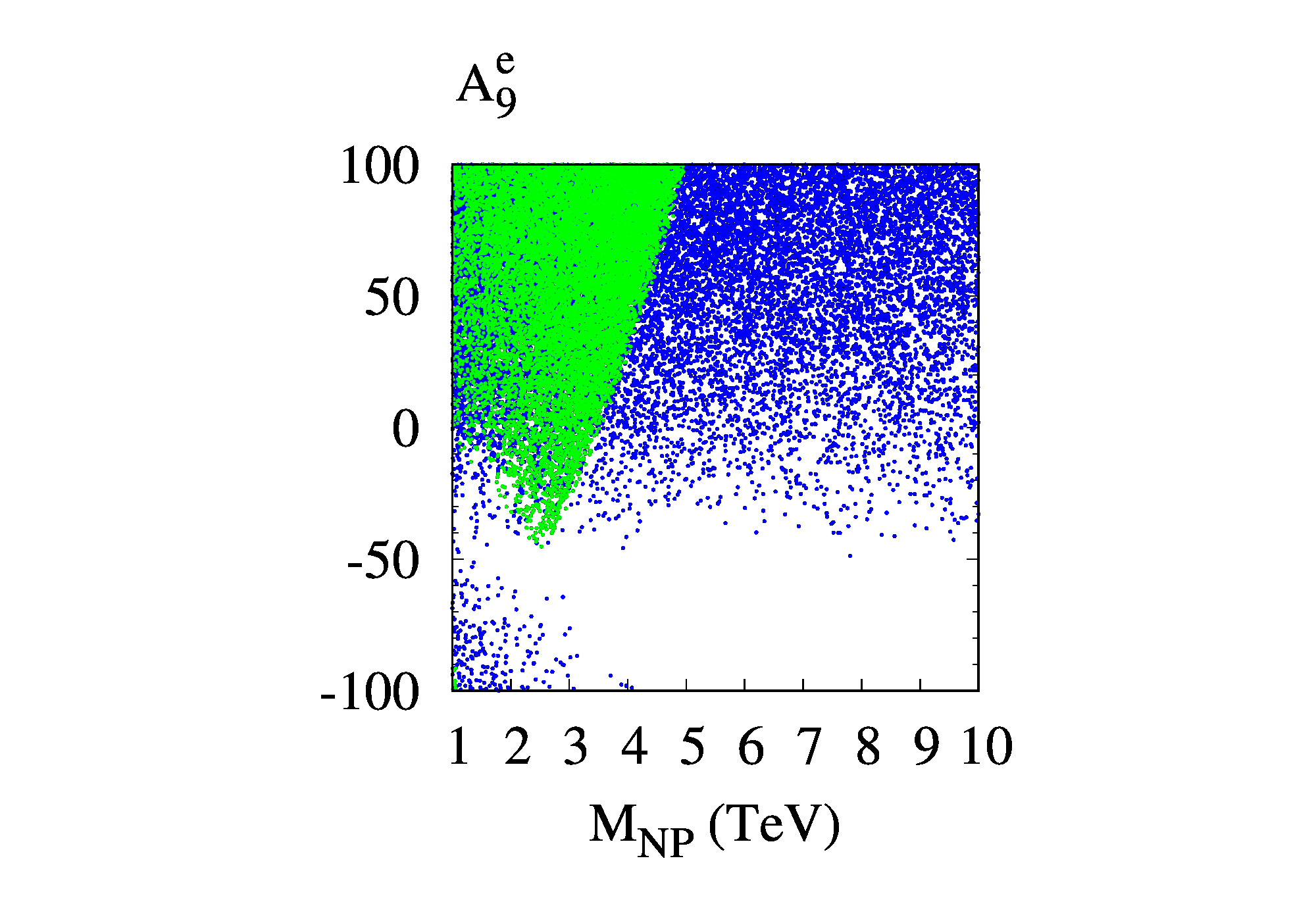} &
\hspace{-2cm}\includegraphics[scale=0.15]{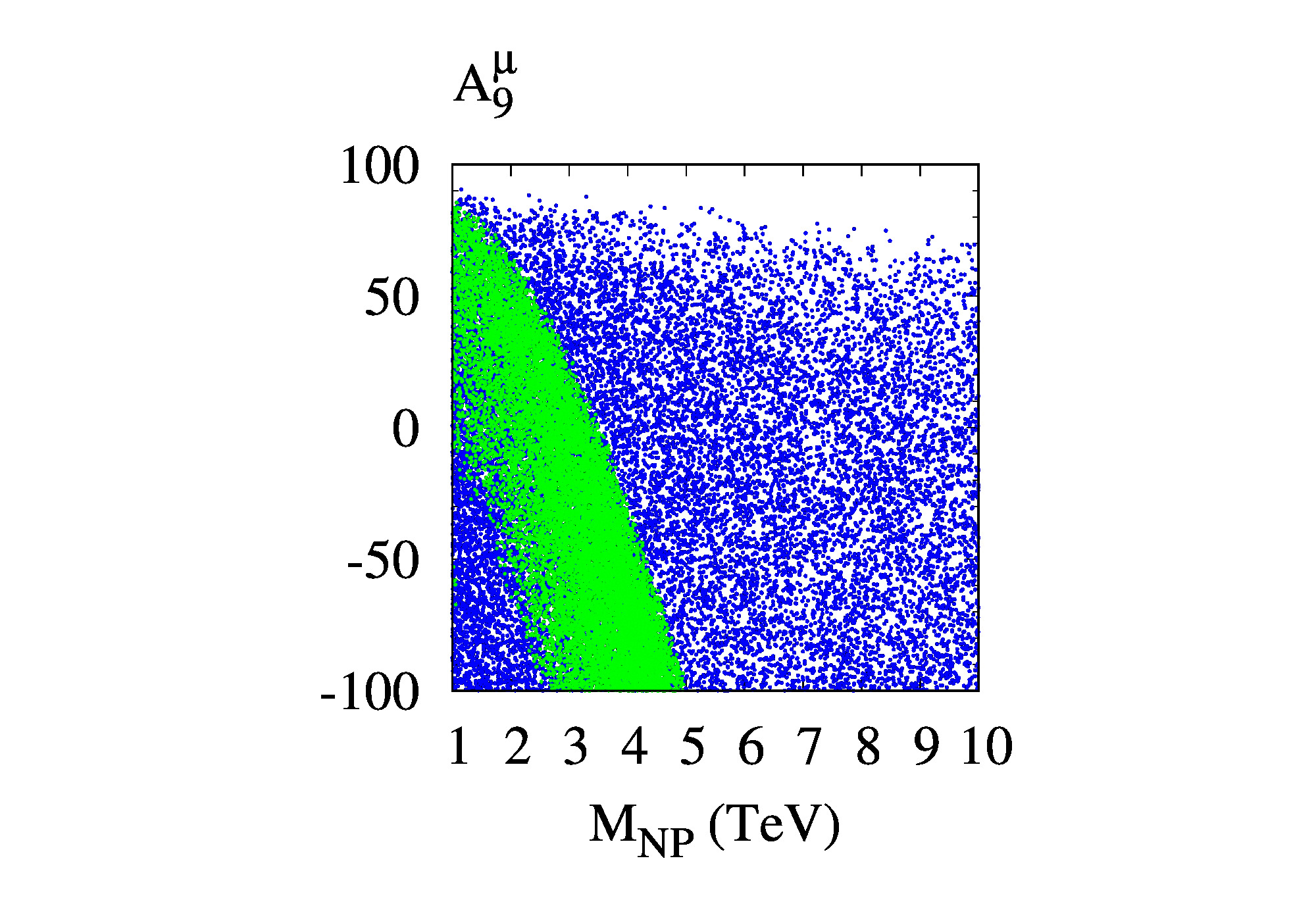} \\
\hspace{-1cm} (a) & \hspace{-2cm} (b) \\
\hspace{-1cm}\includegraphics[scale=0.15]{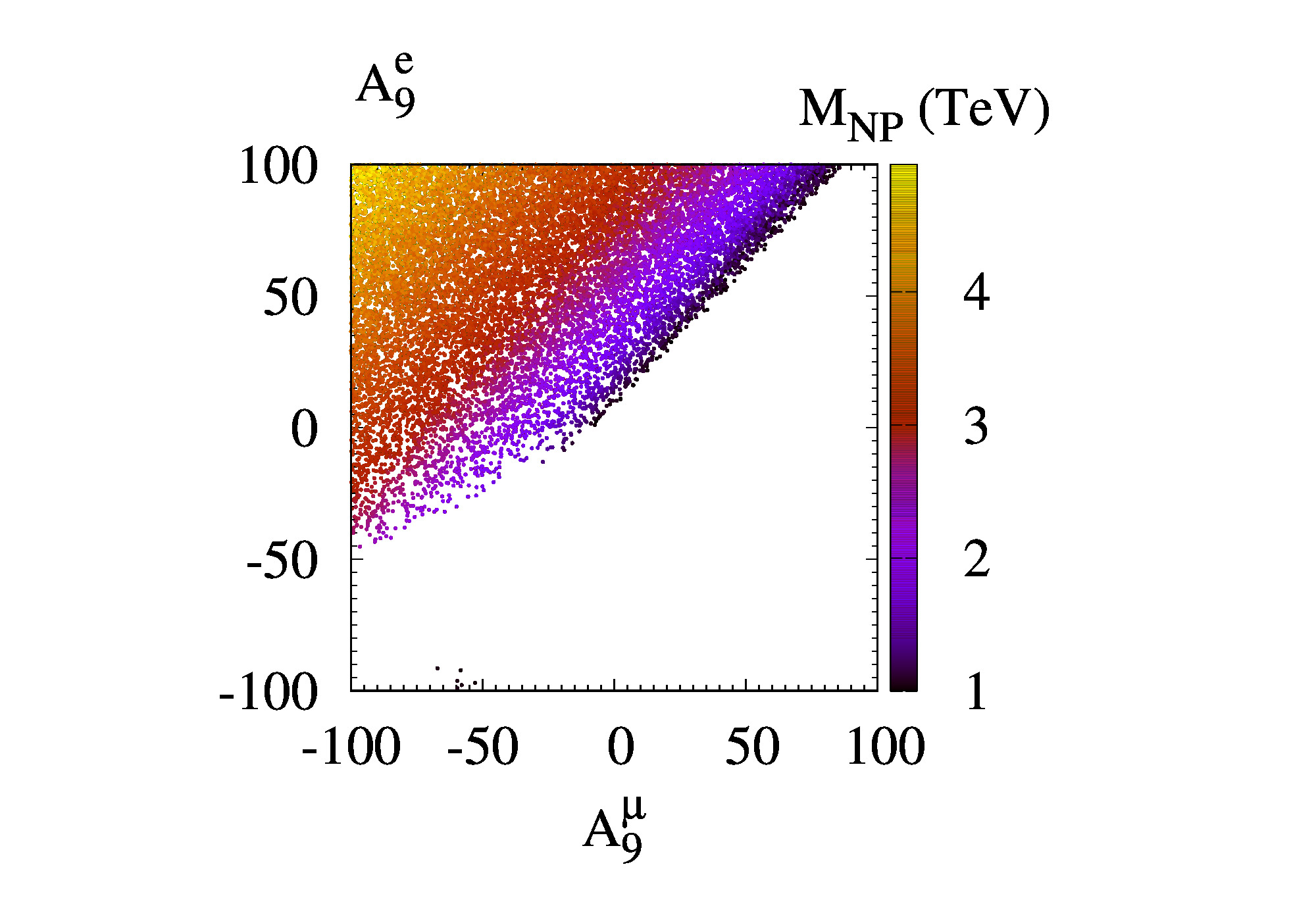} &\\
\hspace{-1cm} (c)
\end{tabular}
\caption{\label{no_cst_00} Allowed regions of (a) $A_9^e$ vs $M_{NP}$ and (b) $A_9^\mu$ vs $M_{NP}$
for free $\alpha$ (blue) and for fixed $\alpha=2$ (green), and of
(c) $A_9^e$ vs $A_9^\mu$ with respect to $M_{NP}$ for $\alpha=2$,
at the $2\sigma$ level without $\Br(\Bs2mumu)$ constraints.}
\end{figure}

\begin{figure}
\begin{tabular}{cc}
\hspace{-2cm}\includegraphics[scale=0.15]{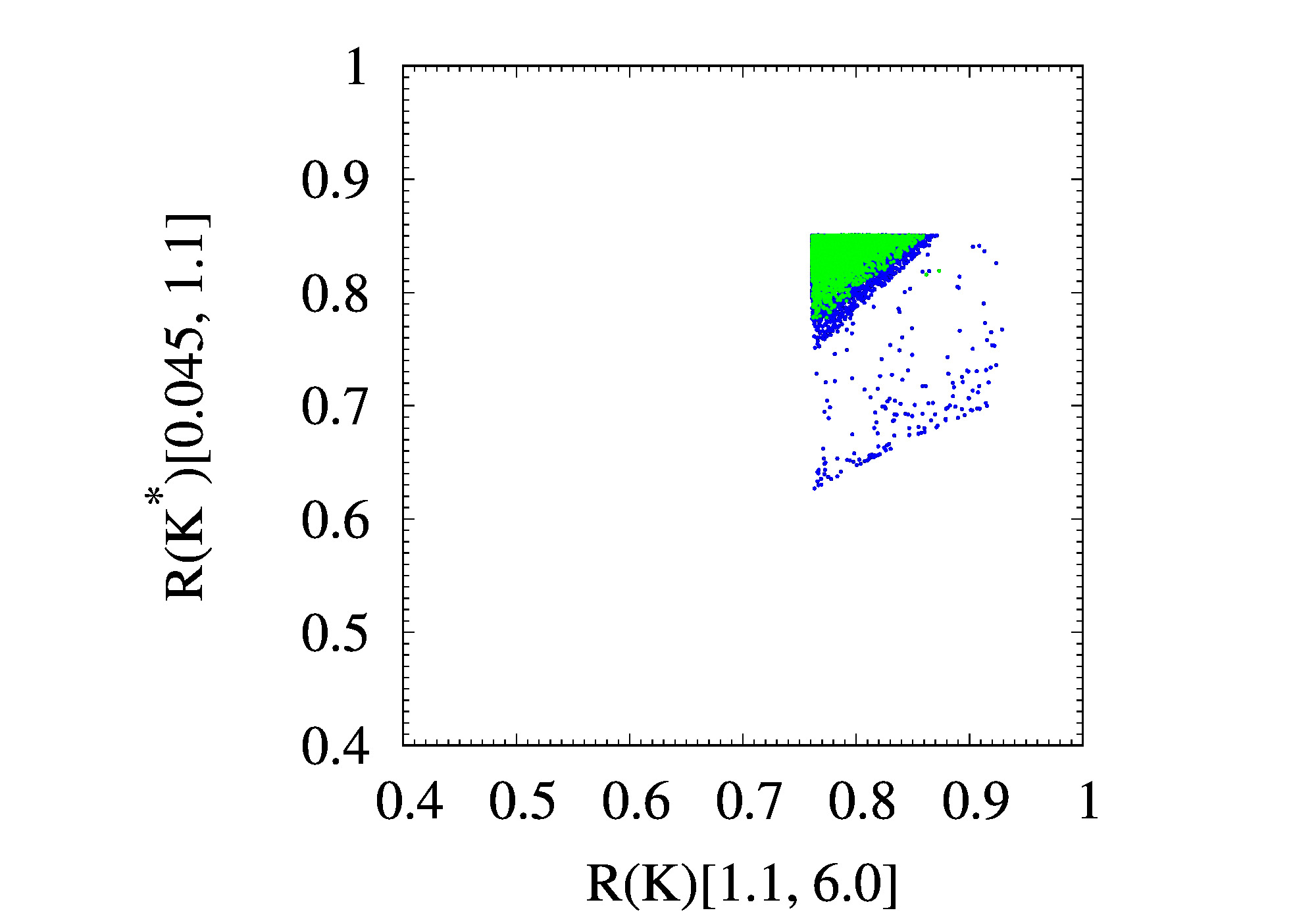} &
\hspace{-2cm}\includegraphics[scale=0.15]{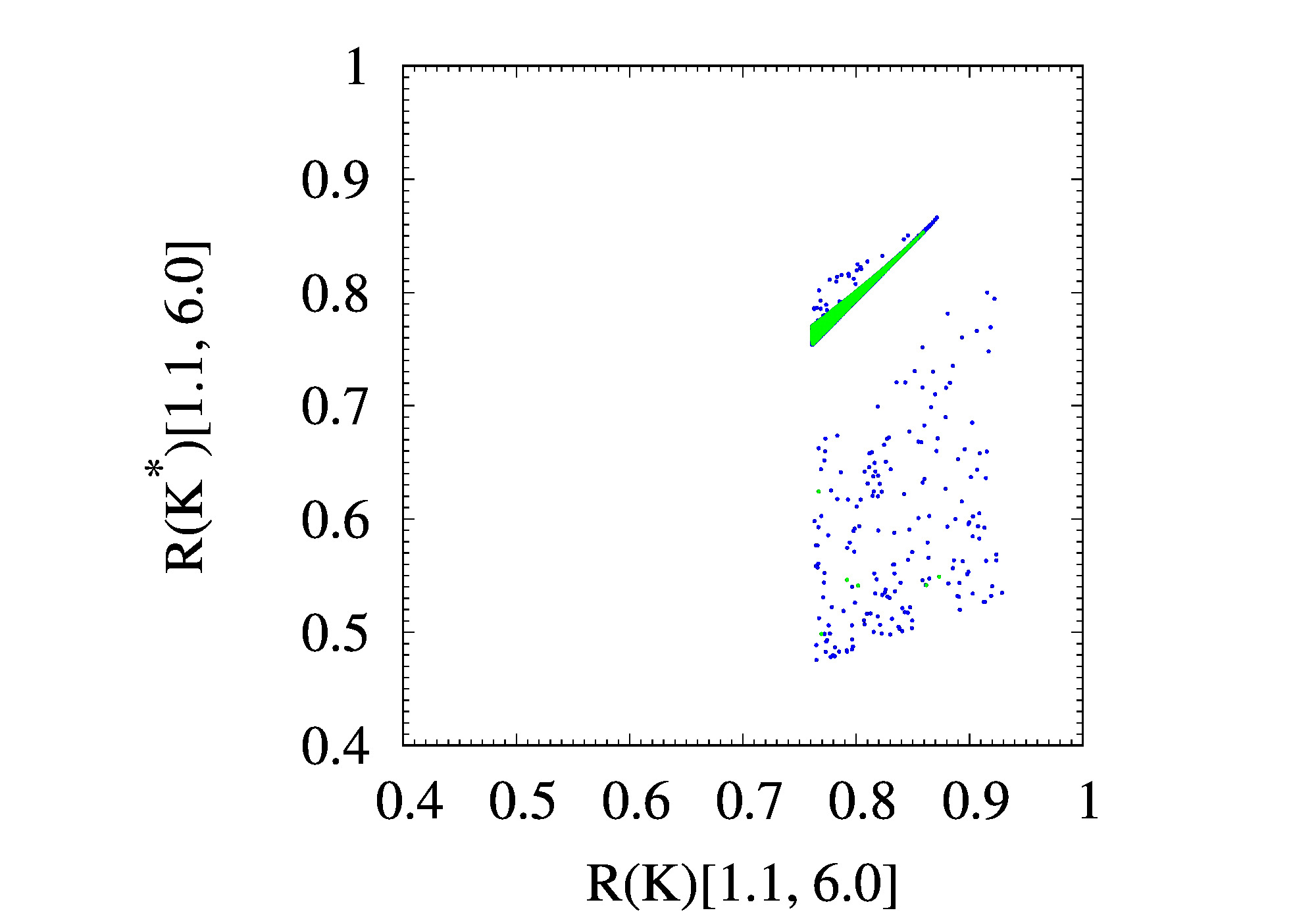} \\
\hspace{-2cm} (a) & \hspace{-2cm} (b) 
\end{tabular}
\caption{\label{no_cst_RKs} Allowed regions of 
(a) $R(K^*)[0.045,1.1]$ vs $R(K)[1.1,6.0]$ and (b) $R(K^*)[1.1,6.0]$ vs $R(K)[1.1,6.0]$
for free $\alpha$ (blue) and for fixed $\alpha=2$ (green) 
at the $2\sigma$ level without $\Br(\Bs2mumu)$ constraints.}
\end{figure}

%
In Fig.\ \ref{no_cst_00} we show $A_9^{e,\mu}$ vs $M_{NP}$ for free $\alpha$ (blue) and for fixed $\alpha=2$ (green).
As expected the parameter spaces shrink a lot for fixed $\alpha=2$.
In our choice of parameter space, only $M_{NP}\lesssim 5~{\rm TeV}$ is allowed when $\alpha=2$..
In Fig.\ \ref{no_cst_00} (c) possible $A_9^{e,\mu}$ are shown with respect to $M_{NP}$ for $\alpha=2$.
It is expected that larger $M_{NP}$ is possible for positively large $A_9^e$ and negatively large $A_9^\mu$.
Figure \ref{no_cst_RKs} depicts the allowed regions of $R(\Ks)$ for fixed $\alpha=2$.
%
\par
\begin{figure}
\begin{tabular}{cc}
\hspace{-1cm}\includegraphics[scale=0.15]{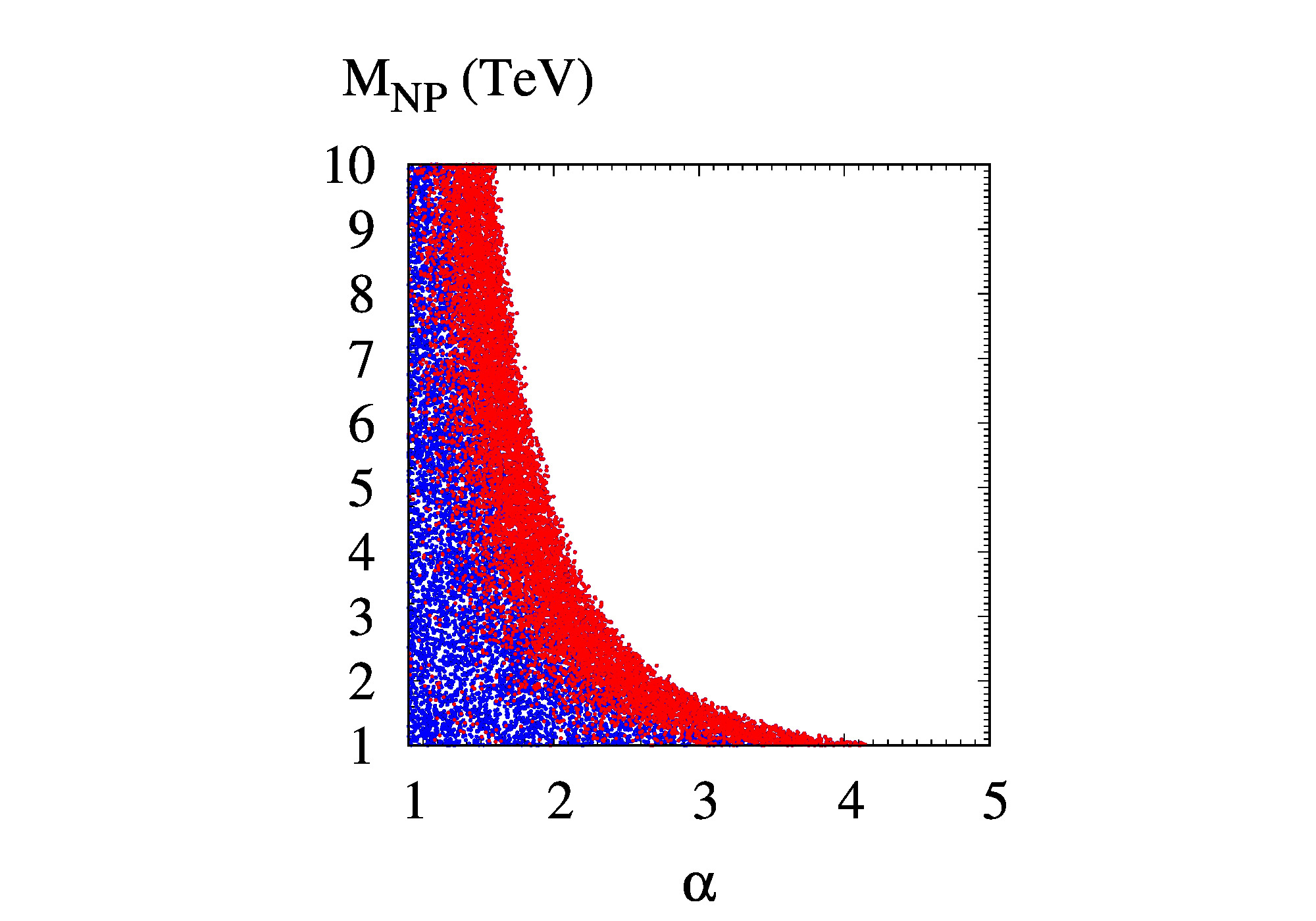} &
\hspace{-2cm}\includegraphics[scale=0.15]{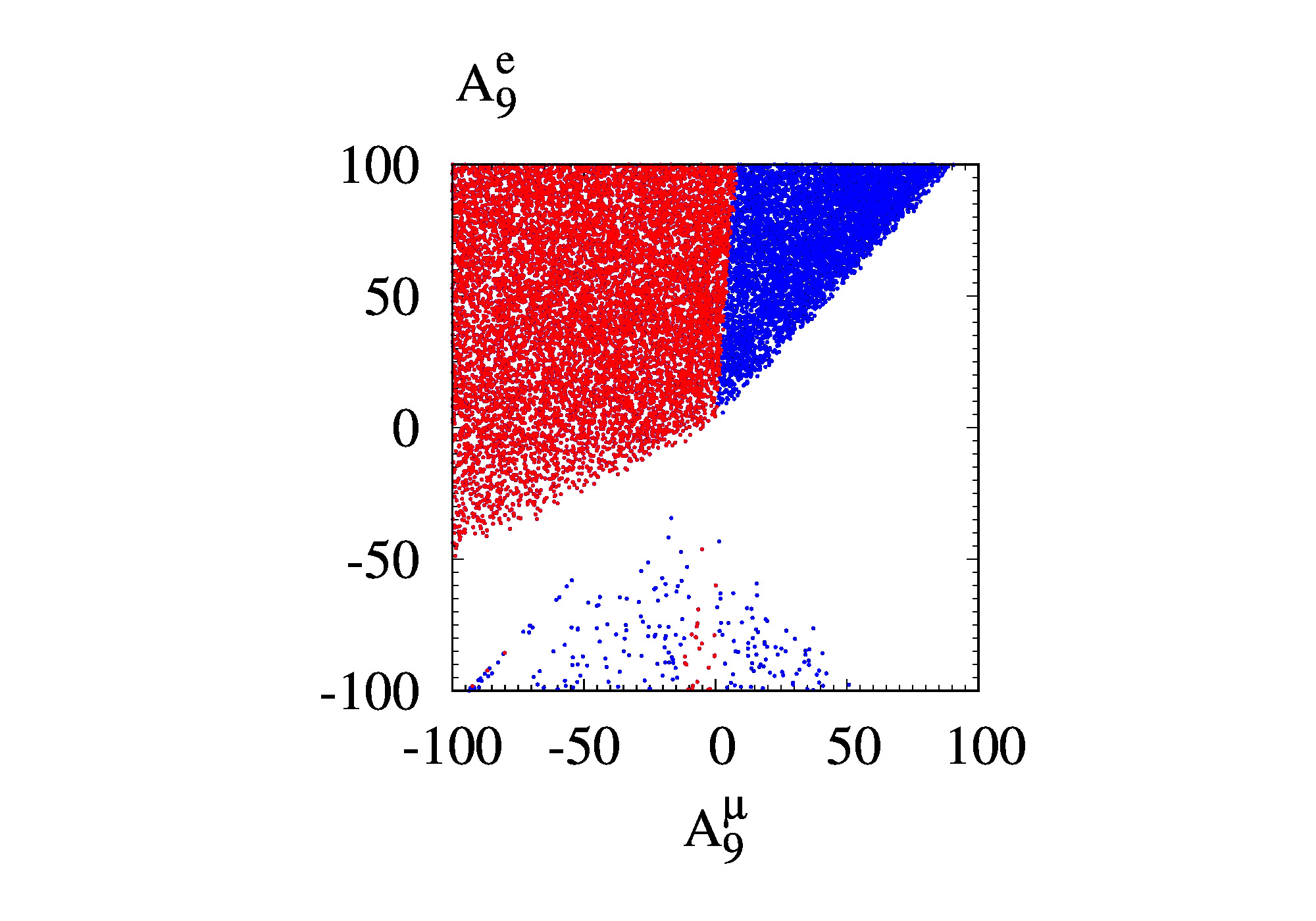}\\
\hspace{-1cm} (a) & \hspace{-2cm} (b) \\
\hspace{-1cm}\includegraphics[scale=0.15]{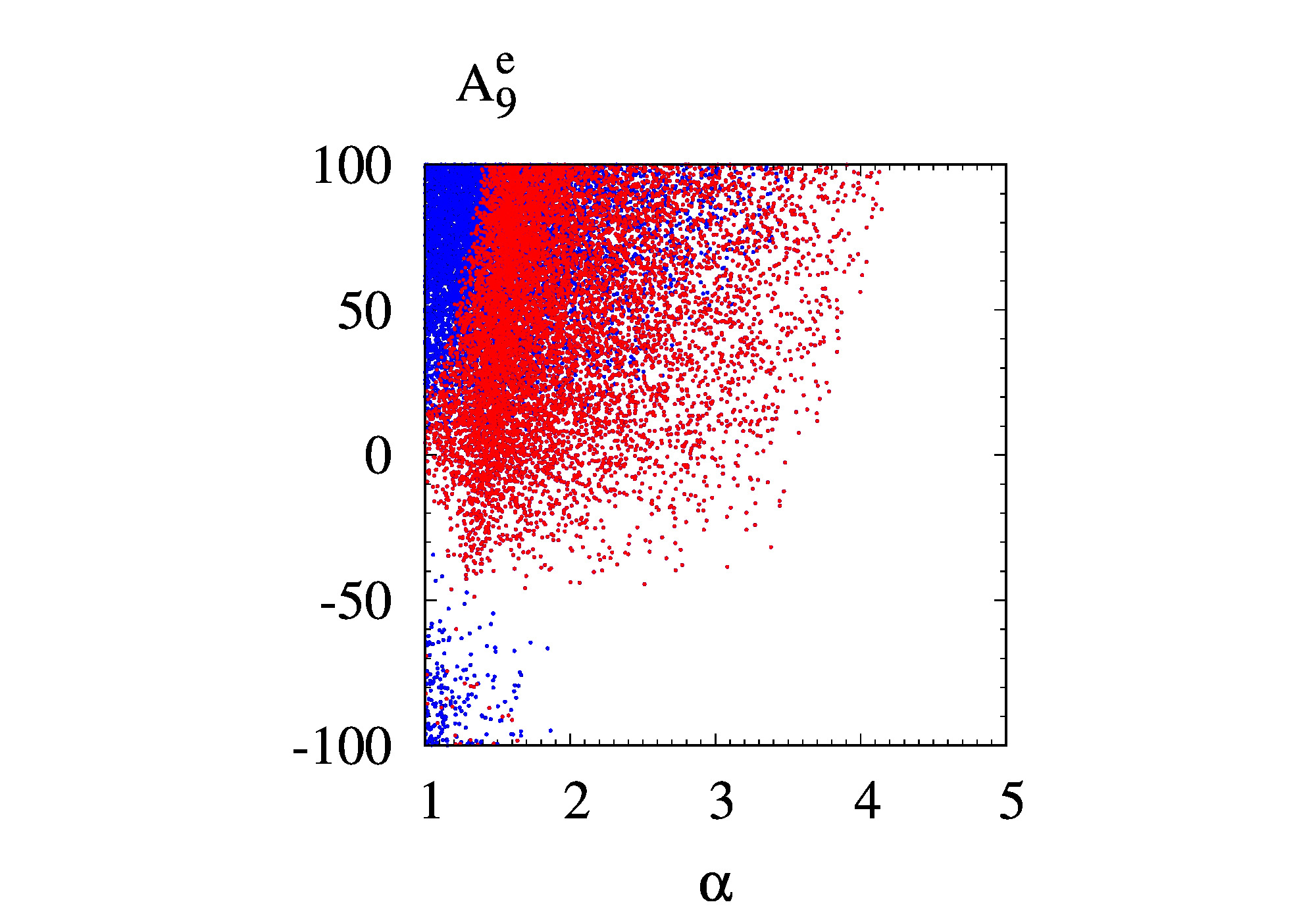} &
\hspace{-2cm}\includegraphics[scale=0.15]{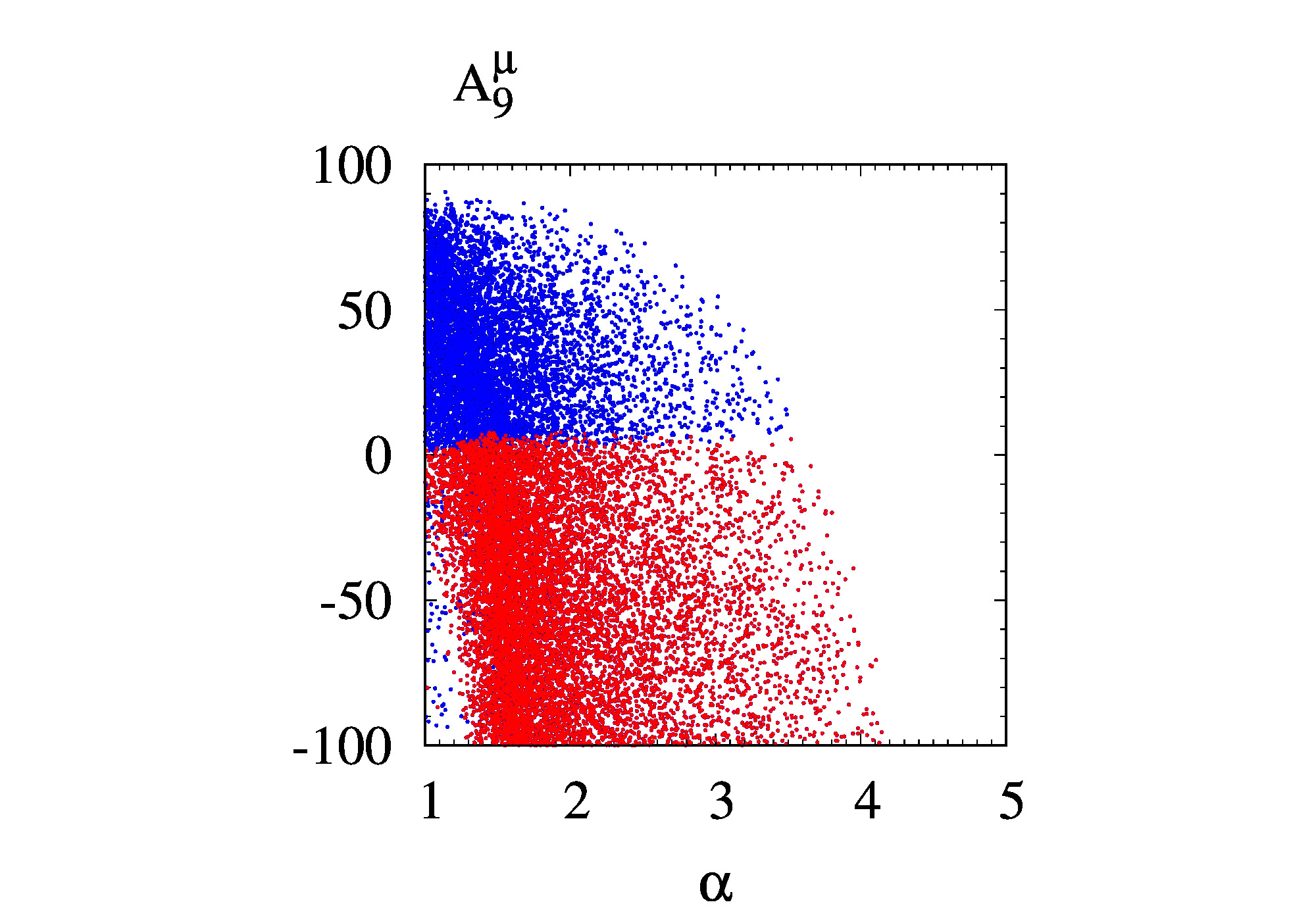} \\ 
\hspace{-1cm} (c) & \hspace{-2cm} (d)
\end{tabular}
\caption{\label{F_cst} Allowed regions of (a) $M_{NP}$ vs $\alpha$, (b) $A_9^e$ vs $A_9^\mu$, 
(c) $A_9^e$ vs $\alpha$,
and (d) $A_9^\mu$ vs $\alpha$ at the $2\sigma$ level.
Constraints from $\Bs2mumu$ are imposed for red dots and not for blue dots.}
\end{figure}

Now we impose the constraints of $\Br(\Bs2mumu)$.
The results are shown in Fig.\ \ref{F_cst}.
We compared with the situation where $\Br(\Bs2mumu)$ is not considered.
As in Fig.\ \ref{F_cst} (a) smaller $\alpha$ and smaller $M_{NP}$ are disfavored.
Such a combination of $\alpha$ and $M_{NP}$ tends to enhance NP effects, which is not compatible with $\Br(\Bs2mumu)$.
Also, Regions of large $A_9^\mu$ and large $A_9^e$ (right upper part of Fig.\ \ref{F_cst} (b)) are avoided.
Just only for $R(\Ks)$, large $A_9^e$ effects would be compensated by large $A_9^\mu$.
But this is not possible when we include $\Br(\Bs2mumu)$ constraints.
The most remarkable result is that $\Br(\Bs2mumu)$ favors mostly negative values of $A_9^\mu$ as in Fig.\ \ref{F_cst} (b).
This is expected from Eqs.\ (\ref{BrBsmumuSM}) and (\ref{BrBsmumuEXP}).
Figures \ref{F_cst} (c) and (d) depict the allowed regions of $A_9^{e,\mu}$ vs $\alpha$.
Small values of $\alpha$ near to unity favor small values of $|A_9^{e,\mu}|$ as expected.
\par
\begin{figure}
\begin{tabular}{cc}
\hspace{-1cm}\includegraphics[scale=0.15]{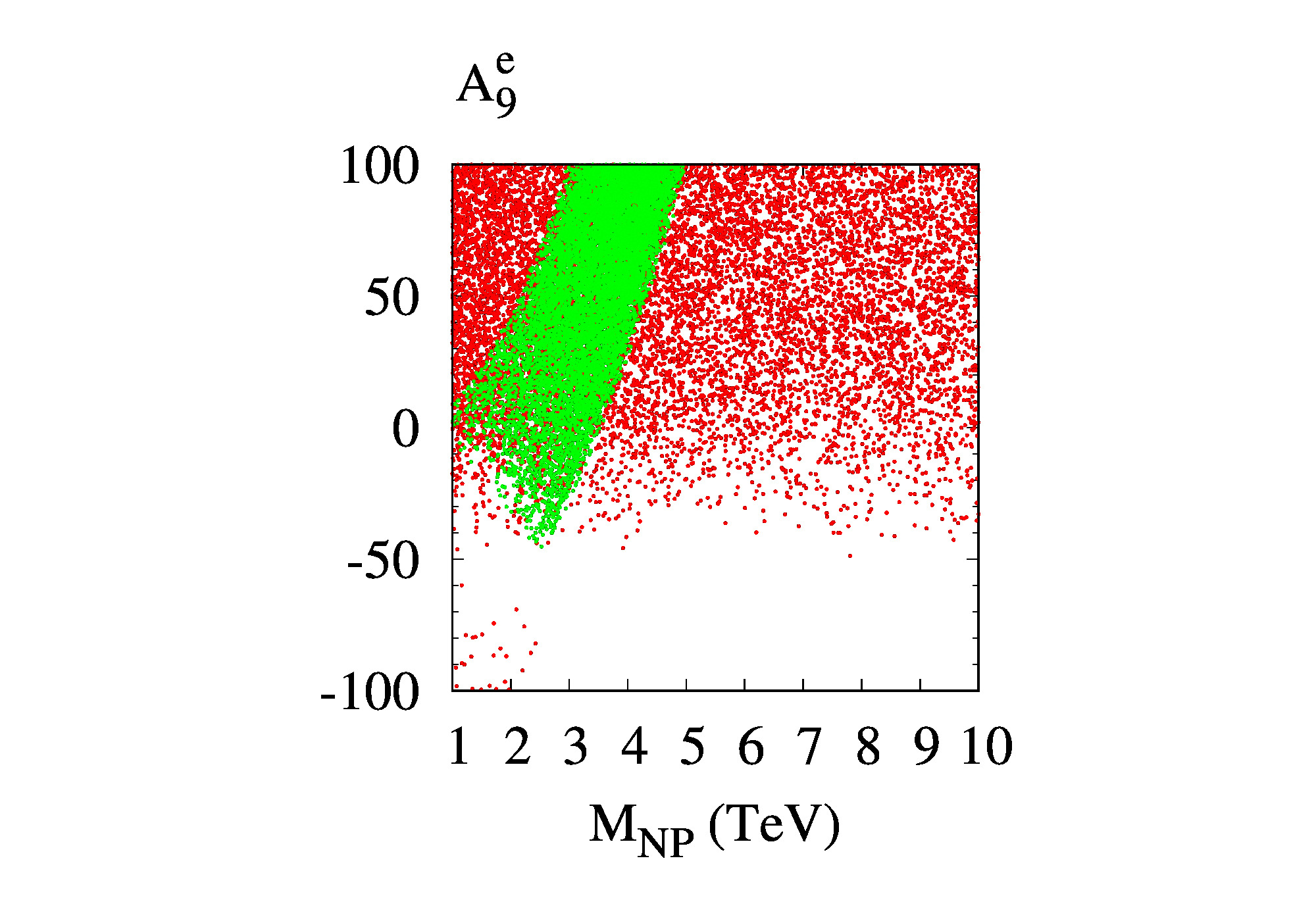} &
\hspace{-2cm}\includegraphics[scale=0.15]{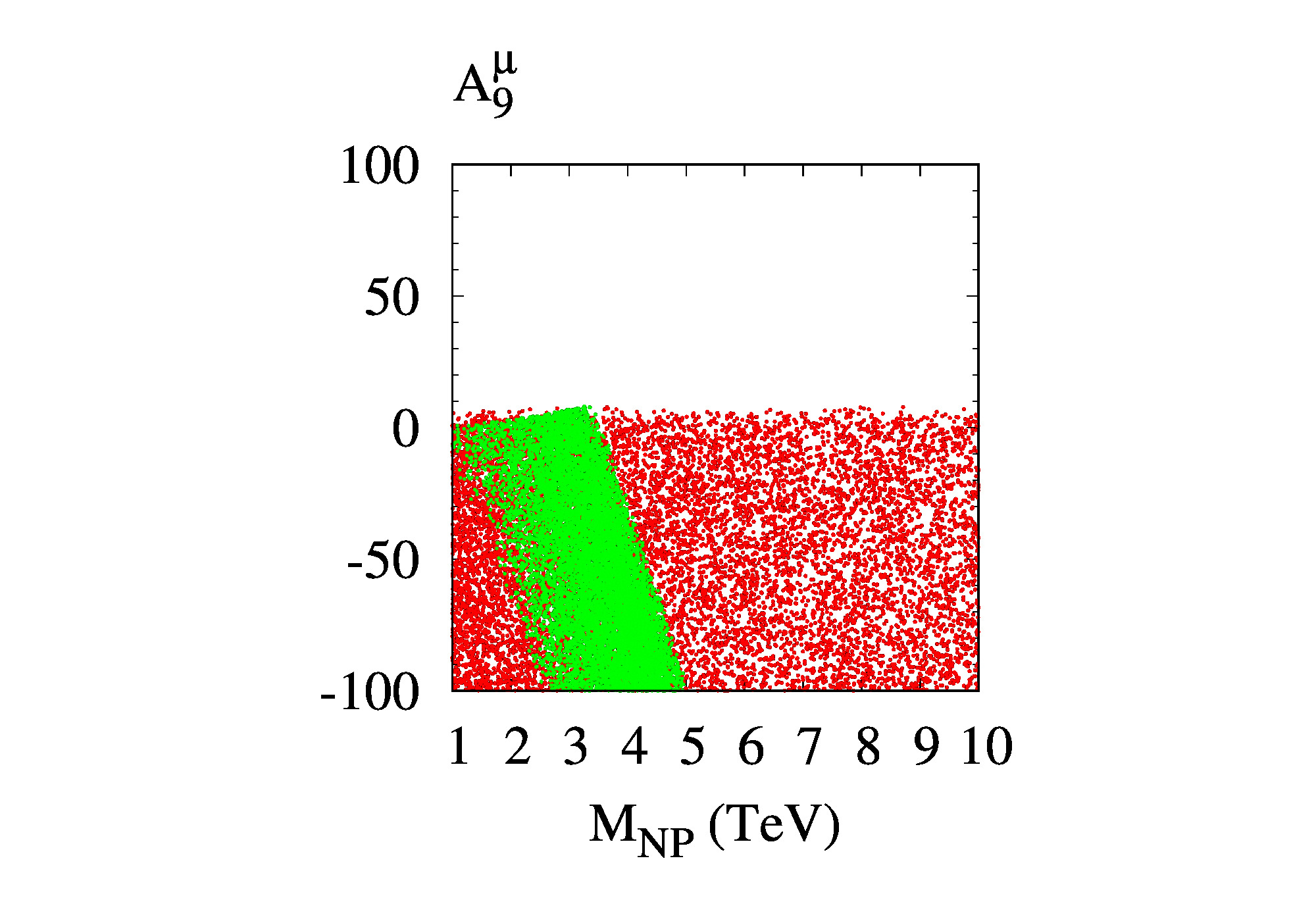} \\
\hspace{-1cm} (a) & \hspace{-2cm} (b) \\
\hspace{-1cm}\includegraphics[scale=0.15]{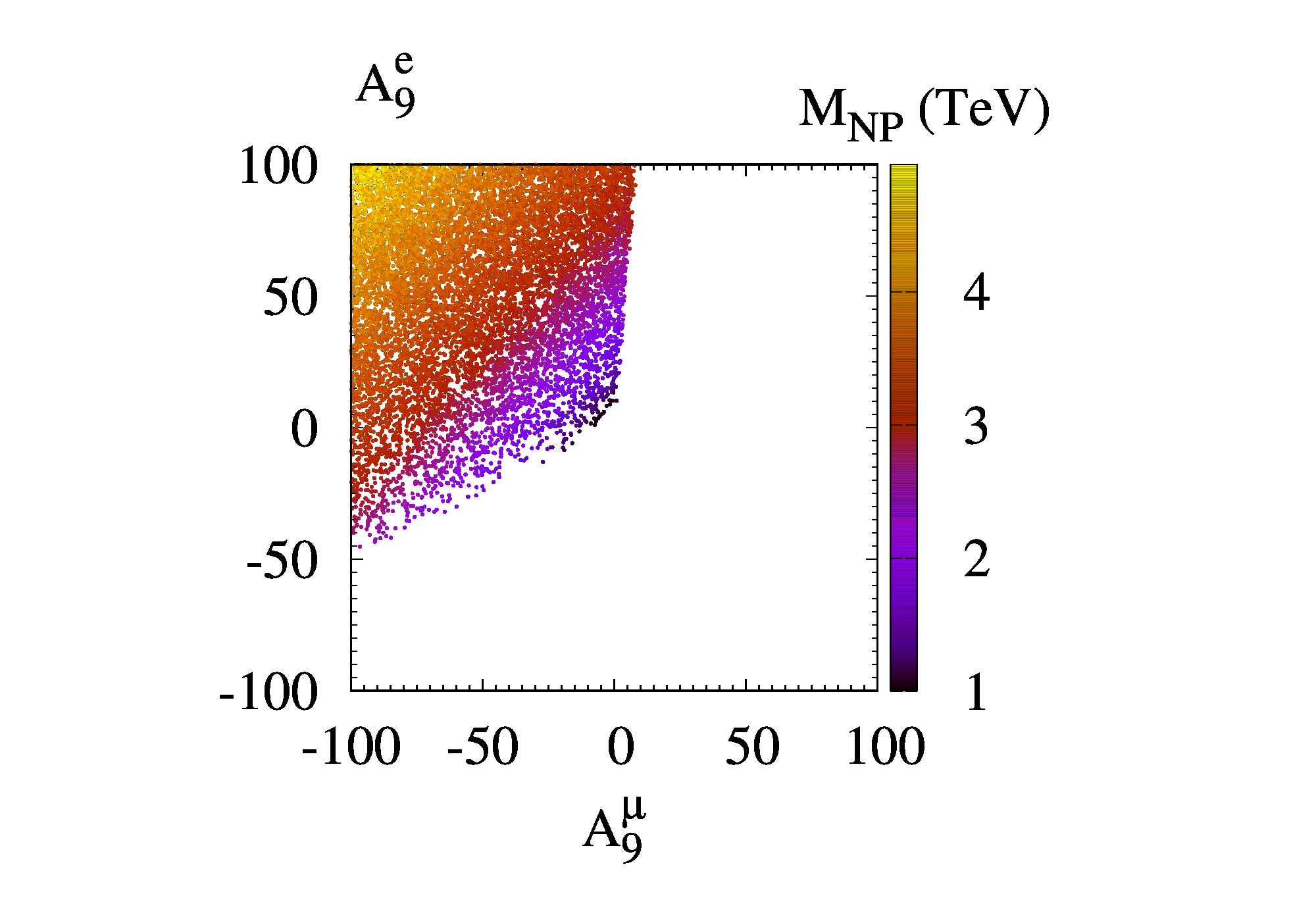} &\\
\hspace{-1cm} (c)
\end{tabular}
\caption{\label{cst_00} Allowed regions of (a) $A_9^e$ vs $M_{NP}$ and (b) $A_9^\mu$ vs $M_{NP}$
for free $\alpha$ (red) and for fixed $\alpha=2$ (green), and of
(c) $A_9^e$ vs $A_9^\mu$ with respect to $M_{NP}$ for $\alpha=2$,
at the $2\sigma$ level with $\Br(\Bs2mumu)$ constraints.}
\end{figure}

Figure \ref{cst_00} shows allowed regions of $A_9^{e,\mu}$ and $M_{NP}$.
As one can see in Figs.\ \ref{cst_00} (a) and (b), possible NP scale for fixed $\alpha=2$ is restricted to
$M_{NP}\lesssim 5~{\rm TeV}$ in our parameter space.
This was already true when $\Br(\Bs2mumu)$ is not considered, 
so the constraint does not restrict the allowed range of $M_{NP}$ so much.
Figure \ref{cst_00} (a) tells that $A_9^e=0$ is also allowed, but for $A_9^e\ne 0$ positive values are more favored.
As seen in Figs.\ \ref{cst_00} (a) and (b), for small $M_{NP}\lesssim 2~{\rm TeV}$ 
large $|A_9^{e,\mu}|$ are not allowed for fixed $\alpha=2$.
For that region NP effects become too large.
Larger $|A_9^{e,\mu}|$ are possible for larger $M_{NP}$, as in Fig.\ \ref{cst_00} (c).
\par
\begin{figure}
\begin{tabular}{cc}
\hspace{-1cm}\includegraphics[scale=0.15]{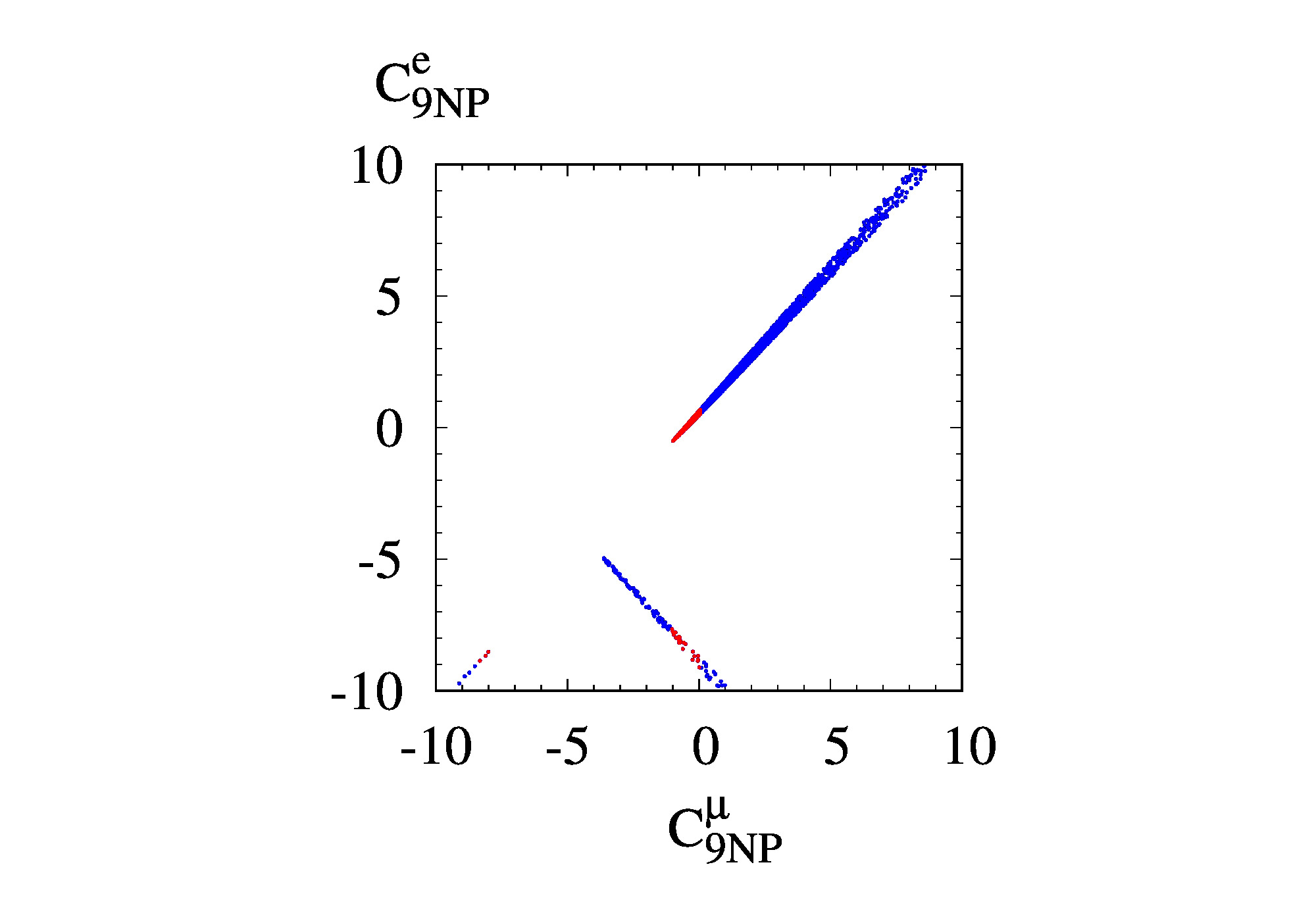} &
\hspace{-2cm}\includegraphics[scale=0.15]{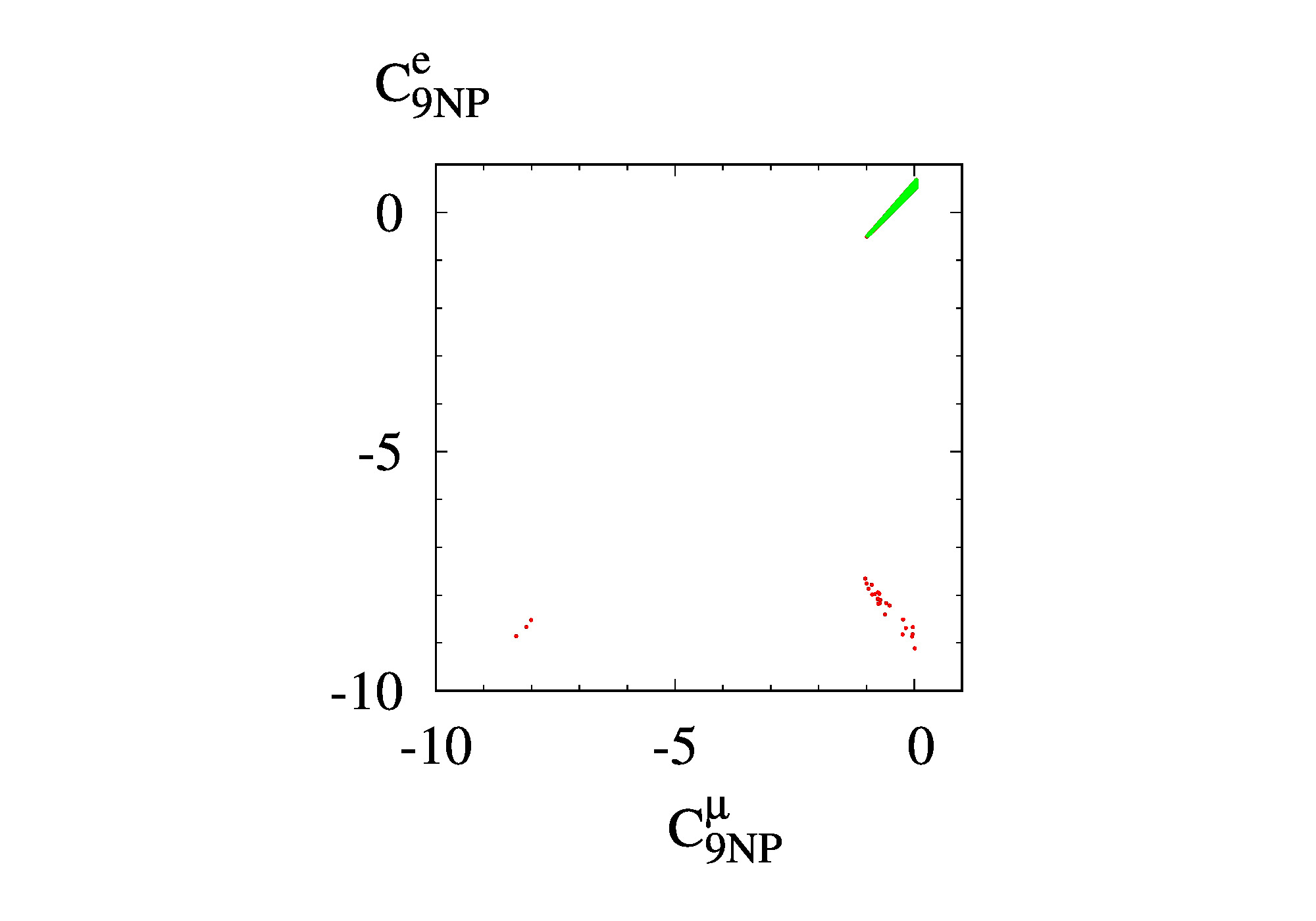} \\
\hspace{-1cm} (a) & \hspace{-2cm} (b) \\
\hspace{-1cm}\includegraphics[scale=0.15]{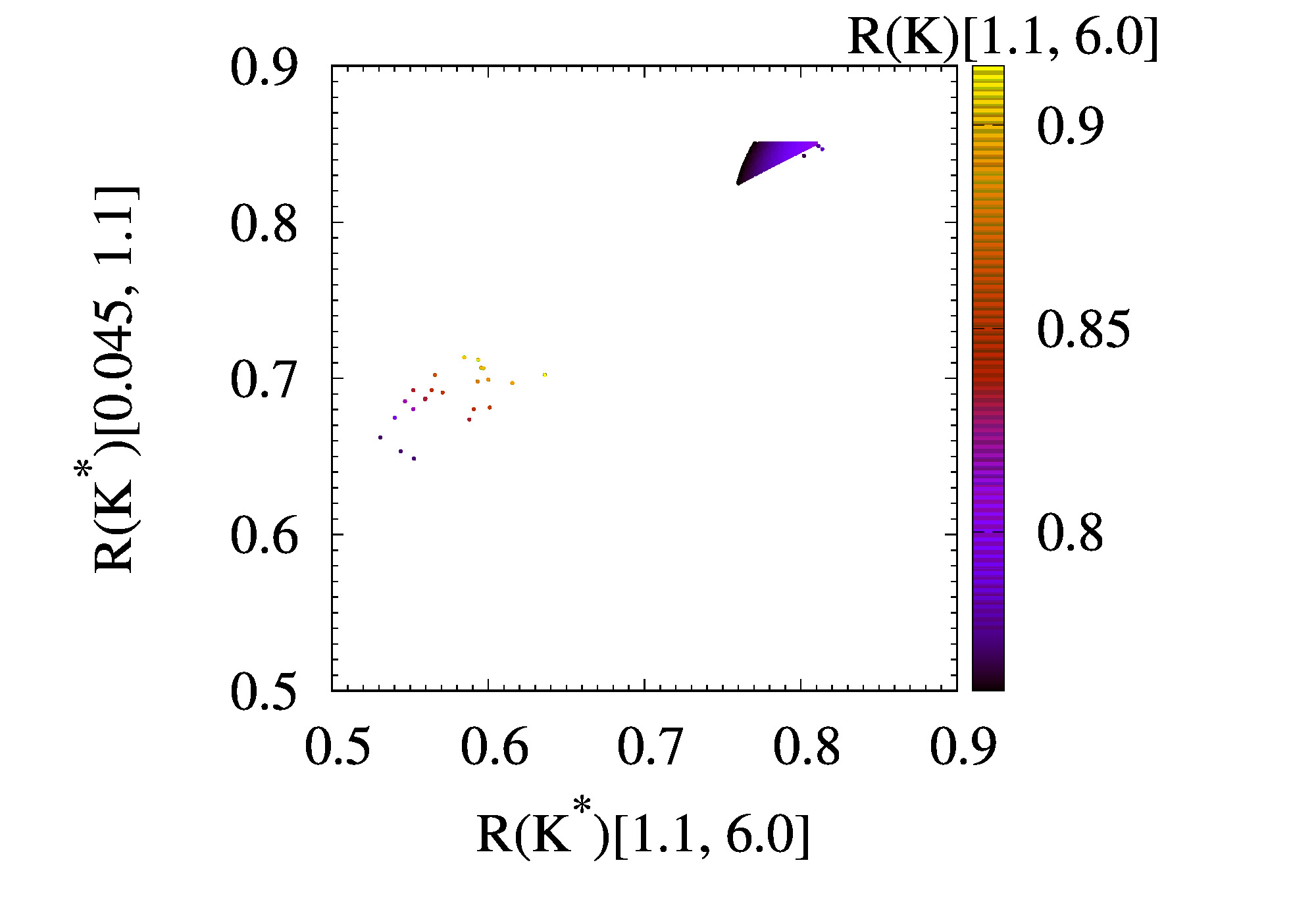} &\\ 
\hspace{-1cm} (c) &
\end{tabular}
\caption{\label{F_cst2} Allowed regions (at the $2\sigma$ level) of 
(a) $C_{9NP}^e$ vs $C_{9NP}^\mu$ with (red) and without (blue) $\Br(\Bs2mumu)$ for free $\alpha$,
(b) $C_{9NP}^e$ vs $C_{9NP}^\mu$ for free $\alpha$ (red) and fixed $\alpha=2$ (green),
and (c) $R(K^*)[0.045,1.1]$ vs $ R(K^*)[1.1,6.0]$ with respect to $R(K)[1.1,6.0]$ with free $\alpha$.
Constraints from $\Bs2mumu$ are imposed in (b) and (c).}
\end{figure}

In Fig.\ \ref{F_cst2} we show the Wilson coefficients and $R(\Ks)$.
It is interesting that there exist three regions of $A_9^{e,\mu}$.
Basically they are some kind of transformation of Fig.\ \ref{F_cst} (b).
From Fig.\ \ref{cst_00} (c), only the left upper part of $A_9^e$-$A_9^\mu$ is allowed when $\alpha=2$, 
which corresponds to $|C_{9N}^{e,\mu}|\lesssim 1$ region of Fig.\ \ref{F_cst2} (b) (green)
and large $R(K^*)$ region (right upper part) of Fig.\ \ref{F_cst2} (c).
Our results for $-1\lesssim C_{9NP}^\mu\lesssim 0$ is compatible with other works \cite{Geng2103}.
\par
Note that $A_9^{e,\mu}$ contains NP couplings to both leptons and quarks.
For a specific model of NP one should consider further constraints for the quark sector by, e.g., $B_s$-$\Bbar_s$ mixing 
and for the leptonic sector by, e.g., lepton anomalous magnetic moments.
If $C_{9NP}$ and $C_{10NP}$ are independent each other, then the constraint from $\Br(\Bs2mumu)$ would be alleviated.
\section{Conclusions}
In conclusion we investigated the $B$ anomaly of $R(\Ks)$ in a model-independent way.
The Wilson coefficients are parameterized by the NP mass scale $M_{NP}$, the power of it $\alpha$, 
and the corresponding coefficients $A_9^{e,\mu}$.
Our choice of parameter windows is very reasonable to result in $\sim\calO(1)$ Wilson coefficients,
which is consistent with other researches.
Our framework can naturally include the constraint from $\Br(\Bs2mumu)$ which is very important for
$b\to s\ell^+\ell^-$ processes.
Without assuming any specific NP models, we could understand general features of the $R(\Ks)$ puzzle 
in model-independent way with our new parametrization.
Especially Fig.\ \ref{F_cst} (b) shows the origin of the lepton-universality violation in $R(\Ks)$.
In our analysis the NP effects on electronic sector might be zero but nonzero case could explain a lot about $R(\Ks)$.
Figure \ref{cst_00} tells possible NP scale up to $\sim 5~{\rm TeV}$ in our parameter window.
We also found that allowed values of $C_{9NP}^{e,\mu}$ are distributed in three regions
and for $\alpha=2$, $C_{9NP}^\mu$ is compatible with other results.
Further detailed constraints should be added in a specific NP scenarios.



\begin{thebibliography}{99}
\bibitem{LHCb2103}
R.~Aaij \textit{et al.} [LHCb],
[arXiv:2103.11769 [hep-ex]].
\bibitem{LHCb1705}
R.~Aaij \textit{et al.} [LHCb],
JHEP \textbf{08} (2017), 055.
\bibitem{Geng2103}
L.~S.~Geng, B.~Grinstein, S.~J\"ager, S.~Y.~Li, J.~Martin Camalich and R.~X.~Shi,
[arXiv:2103.12738 [hep-ph]].
\bibitem{Hiller0310}
G.~Hiller and F.~Kruger,
Phys. Rev. D \textbf{69} (2004), 074020.
\bibitem{Bobeth0709}
C.~Bobeth, G.~Hiller and G.~Piranishvili,
JHEP \textbf{12} (2007), 040.

\bibitem{Geng1704}
L.~S.~Geng, B.~Grinstein, S.~J\"ager, J.~Martin Camalich, X.~L.~Ren and R.~X.~Shi,
Phys. Rev. D \textbf{96} (2017) no.9, 093006.
\bibitem{Hiller1408}
G.~Hiller and M.~Schmaltz,
Phys. Rev. D \textbf{90} (2014), 054014.
\bibitem{Dorsner1603}
I.~Dor\v{s}ner, S.~Fajfer, A.~Greljo, J.~F.~Kamenik and N.~Ko\v{s}nik,
Phys. Rept. \textbf{641} (2016), 1-68.
\bibitem{Bauer1511}
M.~Bauer and M.~Neubert,
Phys. Rev. Lett. \textbf{116} (2016) no.14, 141802.
\bibitem{Chen1703}
C.~H.~Chen, T.~Nomura and H.~Okada,
Phys. Lett. B \textbf{774} (2017), 456-464.
\bibitem{Crivellin1703}
A.~Crivellin, D.~M\"uller and T.~Ota,
JHEP \textbf{09} (2017), 040.
\bibitem{Calibbi1709}
L.~Calibbi, A.~Crivellin and T.~Li,
Phys. Rev. D \textbf{98} (2018) no.11, 115002.
\bibitem{Blanke1801}
M.~Blanke and A.~Crivellin,
Phys. Rev. Lett. \textbf{121} (2018) no.1, 011801.
\bibitem{Nomura2104}
T.~Nomura and H.~Okada,
[arXiv:2104.03248 [hep-ph]].
\bibitem{Angelescu2103}
A.~Angelescu, D.~Be\v{c}irevi\'c, D.~A.~Faroughy, F.~Jaffredo and O.~Sumensari,
[arXiv:2103.12504 [hep-ph]].
\bibitem{Du2104}
M.~Du, J.~Liang, Z.~Liu and V.~Tran,
[arXiv:2104.05685 [hep-ph]].
\bibitem{Crivellin1501}
A.~Crivellin, G.~D'Ambrosio and J.~Heeck,
Phys. Rev. Lett. \textbf{114} (2015), 151801.
\bibitem{Crivellin1503}
A.~Crivellin, G.~D'Ambrosio and J.~Heeck,
Phys. Rev. D \textbf{91} (2015) no.7, 075006.
\bibitem{Chiang1706}
C.~W.~Chiang, X.~G.~He, J.~Tandean and X.~B.~Yuan,
Phys. Rev. D \textbf{96} (2017) no.11, 115022.
\bibitem{King1706}
S.~F.~King,
JHEP \textbf{08} (2017), 019.
\bibitem{Chivukula1706}
R.~S.~Chivukula, J.~Isaacson, K.~A.~Mohan, D.~Sengupta and E.~H.~Simmons,
Phys. Rev. D \textbf{96} (2017) no.7, 075012.
\bibitem{Cen2104}
J.~Y.~Cen, Y.~Cheng, X.~G.~He and J.~Sun,
[arXiv:2104.05006 [hep-ph]].
\bibitem{Davighi2105}
J.~Davighi,
[arXiv:2105.06918 [hep-ph]].
\bibitem{Hu1612}
Q.~Y.~Hu, X.~Q.~Li and Y.~D.~Yang,
Eur. Phys. J. C \textbf{77} (2017) no.3, 190.
\bibitem{Crivellin1903}
A.~Crivellin, D.~M\"uller and C.~Wiegand,
JHEP \textbf{06} (2019), 119.
\bibitem{Rose1903}
L.~Delle Rose, S.~Khalil, S.~J.~D.~King and S.~Moretti,
Phys. Rev. D \textbf{101} (2020) no.11, 115009.
%
\bibitem{Altmannshofer2002}
W.~Altmannshofer, P.~S.~B.~Dev, A.~Soni and Y.~Sui,
Phys. Rev. D \textbf{102} (2020) no.1, 015031.
%
\bibitem{Georgi}
 H.~Georgi,
  Phys.\ Rev.\ Lett.\  {\bf 98}, 221601 (2007);
Phys.\ Lett.\  B {\bf 650}, 275 (2007).
\bibitem{JPL2106}
J.~P.~Lee,
[arXiv:2106.12795 [hep-ph]].
\bibitem{Alonso14}
R.~Alonso, B.~Grinstein and J.~Martin Camalich,
Phys. Rev. Lett. \textbf{113} (2014), 241802.
\bibitem{Geng17}
L.~S.~Geng, B.~Grinstein, S.~J\"ager, J.~Martin Camalich, X.~L.~Ren and R.~X.~Shi,
Phys. Rev. D \textbf{96} (2017) no.9, 093006.
\bibitem{Geng21}
L.~S.~Geng, B.~Grinstein, S.~J\"ager, S.~Y.~Li, J.~Martin Camalich and R.~X.~Shi,
[arXiv:2103.12738 [hep-ph]].
%
\bibitem{Ali99}
A.~Ali, P.~Ball, L.~T.~Handoko and G.~Hiller,
Phys. Rev. D \textbf{61} (2000), 074024.
\bibitem{Chang2010}
Q.~Chang, X.~Q.~Li and Y.~D.~Yang,
JHEP \textbf{04}, 052 (2010).
\bibitem{Beneke1908}
M.~Beneke, C.~Bobeth and R.~Szafron,
JHEP \textbf{10} (2019), 232.
\end{thebibliography}
\end{document}